\documentclass[12pt]{article}
\usepackage{graphicx}
\usepackage{hyperref}
\usepackage{lineno}

\newcommand\pubnumber{SNSN-323-63}
\newcommand\pubdate{\today}

\def\institute{Bergische Universit\"at Wuppertal, GERMANY}

\def\Title#1{\begin{center} {\Large #1 } \end{center}}
\def\Author#1{\begin{center}{ \sc #1} \end{center}}
\def\Address#1{\begin{center}{ \it #1} \end{center}}

\newcommand\pubblock{\rightline{\begin{tabular}{l} \pubnumber\\
         \pubdate  \end{tabular}}}
\newenvironment{Abstract}{\begin{quotation}  }{\end{quotation}}
\newenvironment{Presented}{\begin{quotation} \begin{center} 
             PRESENTED AT\end{center}\bigskip 
      \begin{center}\begin{large}}{\end{large}\end{center} \end{quotation}}

\def\beq{\begin{equation}}
\def\eeq#1{\label{#1}\end{equation}}
\def\eeqn{\end{equation}}

\def\beqa{\begin{eqnarray}}
\def\eeqa#1{\label{#1}\end{eqnarray}}
\def\eeqan{\end{eqnarray}}

\let\bar=\overbar

\def\Dslash{\not{\hbox{\kern-4pt $D$}}}
\def\dslash{\not{\hbox{\kern-2pt $\del$}}}

\def\msb{{\bar{\ssstyle M \kern -1pt S}}}

\def\stt{$\sigma_{t \bar t}$}
\def\met{E_{\mathrm{T}^{\mathrm{miss}}}}
\def\pb{\, \textrm{pb}}

\newcommand{\ttbar}{$t \bar t$}
\newcommand{\pt}{$p_{\mathrm T}$}
\newcommand{\ifb}{$\mathrm{fb}^{-1}$}

\begin{document}
\begin{titlepage}
\pubblock

\vfill
\Title{Measurements of top-quark pair and single top total cross sections, and in-situ systematic constraints}
\vfill
	\Author{Olga Bessidskaia Bylund \\ on behalf of the ATLAS and CMS Collaborations\footnote{Copyright [2018] CERN for the benefit of the ATLAS and CMS collaborations, CC-BY-4.0 license.}}
\Address{\institute}
\vfill
\begin{Abstract}
Measuring the total cross sections of top-quark pair production and single top processes 
at high precision tests the predictions of the Standard Model and can bring better 
understanding of properties such as the top-quark mass, electroweak couplings, 
lepton universality and proton parton distribution functions.
Some of the recent measurements by the ATLAS and CMS experiments at the Large Hadron Collider 
are outlined in this document. Experimental methods are given particular attention, 
particularly the determination of lepton isolation efficiencies in the ATLAS dilepton 
measurement of top-quark pair production.
\end{Abstract}
\vfill
\begin{Presented}
$13^\mathrm{th}$ International Workshop on Top Quark Physics\\
Durham, UK (videoconference), 14--18 September, 2020
\end{Presented}
\vfill
\end{titlepage}
\def\thefootnote{\fnsymbol{footnote}}
\setcounter{footnote}{0}
\paragraph{Introduction}
Measuring the total production cross sections of top-quark pairs (\ttbar) and in single top-quark production 
tests the Standard Model and can help us to better understand various properties of the top-quark, 
electroweak couplings and the structure of the proton. Some recent results from ATLAS~\cite{atlas} 
and CMS\cite{cms} at 
the LHC are outlined below, with a focus on \textit{in-situ} techniques in the \ttbar\ 
dilepton analysis by ATLAS. 

\vspace{-3mm}
\paragraph*{Top-quark pair production in the dilepton channel (ATLAS)}
Measuring \ttbar\ in the dilepton channel while using innovative experimental techniques, 
the ATLAS experiment has pushed the precision on the measured cross section down to 2.4\%. 
A partial Run 2 dataset of 36.1 \ifb collected at 13 TeV collision energy in 2015-2016 is 
used.
In the selection, an electron-muon pair of opposite-sign charge is required along with jets, of which 
either one or two are $b$-tagged. 
The equations for the observed number of events in the 1 $b$-tag and 2 $b$-tag regions ($N_1$, $N_2$) 
are constructed: 
	\setlength\abovedisplayskip{0.5pt}
        \begin{eqnarray}
		N_1 &=& L \sigma_{t \bar t} \epsilon_{e \mu} 2 \epsilon_b (1 - C_b \epsilon_b) + N_1^{bkg} \label{eq1} \\ 
                N_2 &=& L \sigma_{t \bar t} \epsilon_{e \mu} 2 C_b \epsilon_b^2 + N_2^{bkg} \label{eq2}
	\end{eqnarray} 
	    \setlength\belowdisplayskip{0.5pt}
and solved for \stt\ and the $b$-tagging efficiency $\epsilon_b$. 
The correlation coefficient $C_b$ for reconstructing and tagging the two $b$-jets is determined from 
simulation and $N_{1,2}^{bkg}$ denotes the 
background contribution. 
The efficiency to pass the dilepton selection $\epsilon_{e \mu}$ 
is determined from simulation and then corrected using data; 
it is lower in 2016 than in 2015. 
The contamination from QCD multijet events where leptons fail the isolation requirements 
is estimated using a selection where the cut on the isolation variable 
$|d_0|/\sigma_{d_0}$ has been inverted. 
The isolation efficiencies are corrected, accounting for this contamination, by up to 1\% for electrons 
and 0.4\% for muons, see Fig.~\ref{fig:isolep}. 
\begin{figure}[htb]
\centering
\includegraphics[height=4.8cm]{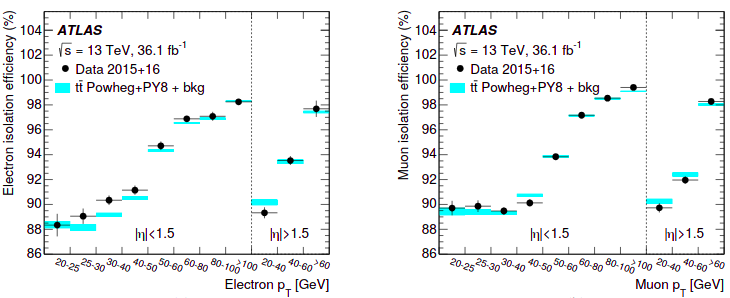}
\caption{The corrections for the lepton isolation efficiencies, binned in 
lepton transverse momentum \pt\ and pseudorapidity $|\eta|$,  for the measurement of \ttbar\ in the 
dilepton channel by the ATLAS experiment \cite{atl_dilep}. The baseline from simulation is 
shown in turquoise and the corrections from data in black.}
\label{fig:isolep}
\end{figure}

Using the BLUE method~\cite{blue1,blue2}, 
the datasets from 2015 and 2016 are combined with weights of 0.49 and 0.51 respectively 
(note that the 2016 dataset is much larger), which 
improves the total sensitivity by 9\% compared to treating them as one dataset. 
Each source of systematic uncertainty is evaluated by changing all
relevant input quantities coherently and re-solving Eqs.~(\ref{eq1}),~(\ref{eq2}). 
The cross section is measured to be: 
\setlength\abovedisplayskip{0.5pt}
	\begin{eqnarray}
		\sigma_{t \bar t}=826.4\pm3.6\mathrm{(stat)}\pm11.5\mathrm{(syst)}\pm15.7\mathrm{(lumi)}\pm1.9\mathrm{(beam)}\pb \, , \nonumber
	\end{eqnarray} 
\setlength\belowdisplayskip{0.5pt}
which is the most precise inclusive measurement of \ttbar\ to date and in agreement with 
\setlength\abovedisplayskip{0.5pt}
\begin{eqnarray}
	\sigma_{t \bar t}^{theo}= 832\pm35\mathrm{(pdf+\alpha_S)}^{+20}_{-29}\mathrm{(scale)} \pb \, . \nonumber \ignorespacesafterend
	\end{eqnarray}
\setlength\belowdisplayskip{0.5pt}
The sensitivity of this measurement is limited by the precision on the luminosity, amounting to an uncertainty 
of 1.9\% on the cross section. The total systematic uncertainty of 1.39\% has the main 
contributions from the normalisation of the $tW$ background and the modelling uncertainties for 
\ttbar. The dilepton analysis is a priori sensitive to modelling uncertainties, but their impact has been 
reduced by measuring the lepton isolation efficiencies \textit{in-situ}: when comparing the predictions from 
different generators, the different predictions for these efficiencies need not be included.

Additionally, the ratio of \ttbar\ to $Z$ production is evaluated to nearly cancel the large dependency on the luminosity and 
constrain proton p.d.fs.  
Ratios of \ttbar\ production at different collision energies and double ratios of \ttbar\ to $Z$ 
production at different energies are also computed, leading to further cancellations.
%

\vspace{-3mm}
\paragraph*{Top-quark pair production in the dilepton channel (CMS)}
In the \ttbar\ cross section measurement in the dilepton channel by CMS~\cite{cms_dilep}, 
in addition to $e^\pm \mu^\mp$, the opposite-sign $ee$ and $\mu \mu$ channels are considered.
35.9 \ifb\ of data collected at 13 TeV collision energy is used for the measurement. 
A total of 28 event categories are defined by splitting by jet and $b$-jet multiplicity in these channels: with 
0, 1 or 2 $b$-tagged jets and 0, 1, 2 or at least 3 additional un-tagged jets (the no $b$-jet 
categories are only used for $e \mu$ events).
A simultaneous template fit using all event categories is performed, fitting the transverse momentum \pt\ of the 
additional jet with highest \pt, if defined, otherwise the event yield is fitted.
A priori, the identification efficiency is known to better precision for muons than for electrons. By using 
the different lepton channels, the electron identification efficiency is constrained to the precision for muons. 
Using categories with different $b$-jet multiplicities allows constraining the efficiency for identifying 
and selecting a $b$-jet. The total cross section is measured to: 
\setlength\abovedisplayskip{0.5pt}
\begin{eqnarray}
	\sigma_{t \bar t} = 803 \pm 2 \mathrm{(stat)} \pm 25 \mathrm{(syst)} \pm 20 \mathrm{(lumi)} \pb \, , \ignorespacesafterend \nonumber
\end{eqnarray} 
\setlength\belowdisplayskip{0.5pt}
in agreement with theory. The main uncertainties relate to the integrated luminosity and lepton identification 
and isolation, with an impact of 2.5\% and 2.0\% respectively. 

\vspace{-3mm}
\paragraph*{Top-quark pair production, lepton+jets (ATLAS)}
The ATLAS experiment has performed a measurement of the \ttbar\ cross section in the lepton+jets channel using 
the full Run 2 dataset (139 \ifb) at 13 TeV \cite{atl_ljets}.
Exactly one charged lepton is required as well as at least four jets. Three event categories are defined, split 
by jet and $b$-jet multiplicity with: 4 jets with 1 $b$-tagged in the SR1 category, 4 jets of which 2 are 
$b$-tagged in SR2 and at least 5 jets of which 2 are $b$-tagged in SR3. 
The aplanarity is used as the fitting variable in SR1, the minimum invariant mass over all lepton-jet pairs in 
SR2 and a distance parameter between the jets in SR3. 
The result is: 
\setlength\abovedisplayskip{0.5pt}
\begin{eqnarray}
	\sigma_{t \bar t} = 830 \pm 0.4 \mathrm{(stat)} \pm 36 \mathrm{(syst)} \pm 14 \mathrm{(lumi)} \pb, \nonumber
\end{eqnarray}
\setlength\belowdisplayskip{0.5pt}
with a precision of 4.6\% and in agreement with the theoretical prediction. This is the most precise 
inclusive \ttbar\ measurement in this channel. The measurement is systematically limited, with the 
main uncertainties originating from parton shower systematics, followed by luminosity  and final 
state radiation.
\vspace{-3mm}
\paragraph*{Top-quark pair production with a hadronic $\tau$ (CMS)}
CMS has measured \ttbar\ production in a selection targeting a hadronically decaying $\tau$ lepton 
and one electron or muon, using 35.9 \ifb\ of data collected at 13 TeV collision energy
~\cite{cms_tau}. This channel corresponds  to around 5\% of all \ttbar\ final states and serves as 
a test of lepton universality. 
One electron or muon, a hadronic $\tau$ candidate of opposite-sign charge and two jets are required, of which at 
least one should be $b$-tagged. The selection is further split into categories by cutting on
a distance parameter between the jets.

The $\tau$ leptons are reconstructed with the hadron-plus-strips (HPS) algorithm. 
In each jet, a 
charged hadron is combined with other nearby charged hadrons or photons and the decay modes of the 
$\tau$ are identified. Electrons and photons are clustered in strips along the bending direction of 
the trajectory, which enhances the identification of $\pi^0$ mesons by taking into account the early 
showering of photons.
A boosted decision tree (BDT) is trained to separate HPS $\tau$ candidates from QCD multijet events. 
The largest background in the measurement comes from \ttbar\ events with a jet misidentified as originating 
from a $\tau$. 
The uncertainty in the efficiency of $\tau$ identification is 5\% and has the highest impact on the precision.
To extract the cross section, the transverse mass $M_T(\ell, \met)$ is fitted 
in a profile likelihood fit, giving: 
\setlength\abovedisplayskip{0.5pt}
\begin{eqnarray}
	\sigma_{t \bar t} = 781 \pm 7 \mathrm{(stat)} \pm 62 \mathrm{(syst)} \pm 20 \mathrm{(lumi)} \pb \, , \nonumber
\end{eqnarray}
\setlength\belowdisplayskip{0.5pt}
which is consistent with the Standard Model prediction. 
This is the first measurement of \ttbar\ production with a $\tau$ lepton in the
final state performed at 13 TeV. 
The ratio to \ttbar\ production in the dilepton channel is computed and found to be in 
agreement with lepton universality. 

\vspace{-3mm}
\paragraph*{Single top production in the $t$-channel (CMS)}
The CMS experiment has measured the single top production cross section in the $t$-channel 
\cite{cms_tch}, using 35.9 \ifb\ of data collected at 13 TeV. This process is 
charge asymmetric due to the valence quark content of the colliding protons.
An electron or muon is required in the selection together with jets. In the signal-enriched category 
exactly one $b$-tagged jet and one additional jet are required. 
The cross sections are measured separately for top and top antiquark production by splitting the events 
into categories by the charge of the lepton and extracting them in a simultaneous fit. 
Their ratio
$R_t = \sigma_{t}/\sigma_{\bar t}$, which is sensitive to proton p.d.fs, is also determined, 
see Fig.~\ref{fig:tch}.
\begin{figure}[htb]
\centering
	\includegraphics[height=2.8cm]{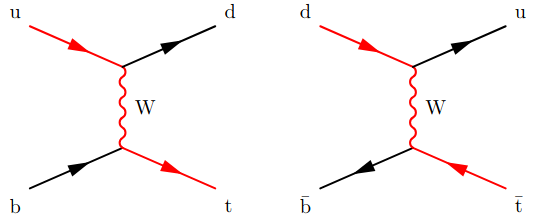} $\quad$
\includegraphics[height=5cm]{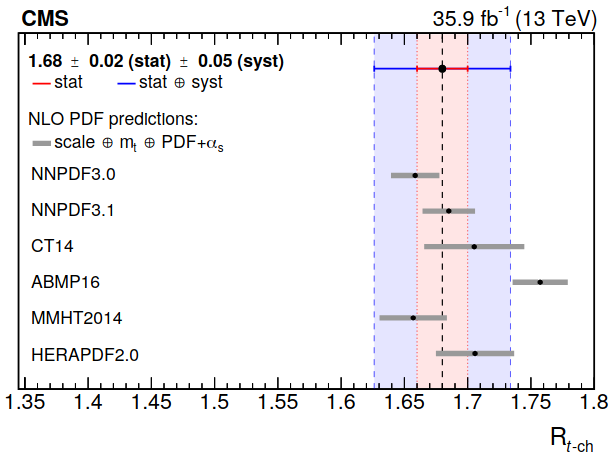}
	\caption{Feynman diagrams illustrating $t$-channel production and the measured value for $R_t$ by the CMS experiment \cite{cms_tch}, compared to different p.d.fs.}
\label{fig:tch}
\end{figure}
BDT algorithms are employed to enhance 
the separation of the signal from the large backgrounds consisting mostly of \ttbar\ and $W$+jets events.
Experimental uncertainties and uncertainties on the background rates and modelling 
enter as nuisance parameter in the profile likelihood fit, which 
allows them to be constrained. The uncertainties relating to signal 
modelling 
are not profiled to 
avoid propagating any constraints on them to the full phase space, where they may not be 
valid.
The cross sections and their ratio are measured to be: 
\setlength\abovedisplayskip{0.5pt}
\begin{eqnarray}
	\sigma_t=&130\pm1\mathrm{(stat)}\pm19\mathrm{(syst)}\pb &   ,  \nonumber \\
	\sigma_{\bar t}=&77\pm1\mathrm{(stat)}\pm12\mathrm{(syst)}\pb &  ,  \quad 
	R_t=1.68\pm0.02\mathrm{(stat)}\pm0.05\mathrm{(syst)} \, . \nonumber
\end{eqnarray}
\setlength\belowdisplayskip{0.5pt}
This is consistent with the predictions from the Standard Model: 
\setlength\abovedisplayskip{0.5pt}
\begin{eqnarray}
	\sigma_t^{theo}&=136.0^{+4.1}_{-2.9}\mathrm{(scale)}\pm3.5\mathrm{(pdf+\alpha_S)}\pb & , \nonumber \\
	\sigma_{\bar{t}}^{theo}&=81.0^{+2.5}_{-1.7}\mathrm{(scale)}\pm3.2\mathrm{(pdf+\alpha_S)}\pb & , \quad R_t=1.68 \, . \nonumber
\end{eqnarray}
\paragraph*{Single top combination and measurements of $\mathbf{tW}$} 

ATLAS and CMS have performed a combination of single top measurements at 7 and 8 TeV \cite{combi} 
using the BLUE method. The results are summarised in Fig.~\ref{fig:combi}.
\begin{figure}[htb]
\centering
	\includegraphics[height=5.4cm]{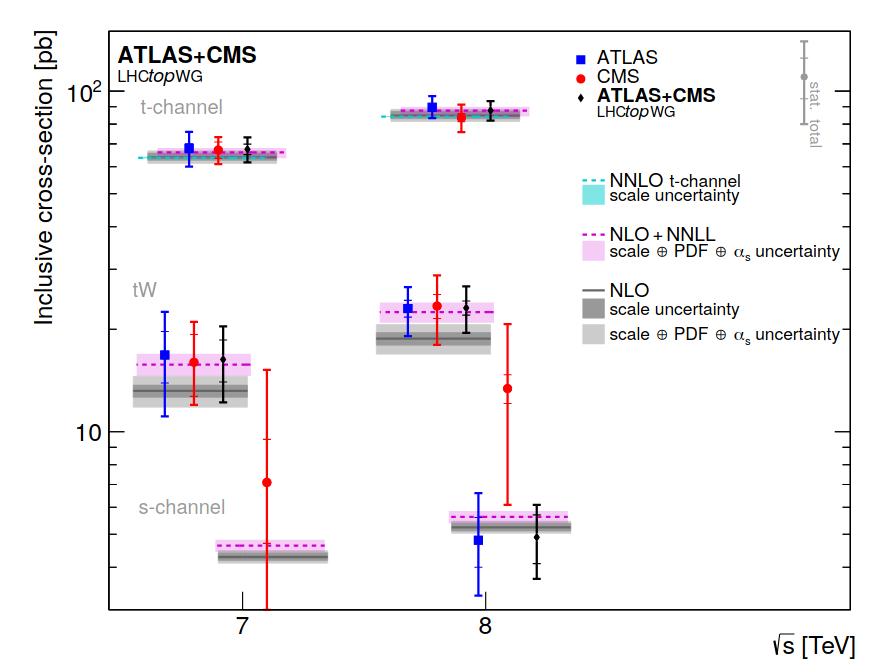}
	\caption{The combination of single top measurements by ATLAS and CMS at 7 and 8 TeV \cite{combi}. The theoretical prediction is shown in gray, the ATLAS result in blue, the CMS result in red and the combination in black.}  
\label{fig:combi}
\end{figure}
Since then, ATLAS also published a measurement of $tW$ production at 8 TeV in the challenging lepton+jets channel \cite{atl_tW}, using 20.2 \ifb of data, obtaining: 
\setlength\abovedisplayskip{0.5pt}
\begin{eqnarray}
	\sigma_{tW} = 26 \pm 7 \pb \, , \nonumber
\end{eqnarray}
with a statistical uncertainty of 4 pb. The result agrees with the theoretical prediction:
\setlength\abovedisplayskip{0.5pt}
\begin{eqnarray}
\sigma_{tW}^{theo} = 22.4 \pm 0.6\mathrm{(scale)} \pm 1.4\mathrm{(pdf)} \pb \, . \nonumber
\end{eqnarray}
\setlength\belowdisplayskip{0.5pt}

\vspace{-4mm}
CMS reported the most precise measurement of $tW$ at 13 TeV \cite{cms_tW} with 35.9 \ifb: 
\setlength\abovedisplayskip{0.5pt}
\begin{eqnarray}
	\sigma_{tW} = 63.1 \pm 1.8 \mathrm{(stat)} \pm 6.4 \mathrm{(syst)} \pm 2.1 \mathrm{(lumi)} \pb \, , \nonumber
\end{eqnarray}
\setlength\belowdisplayskip{0.5pt}
		consistent with 
\setlength\abovedisplayskip{0.5pt}
\begin{eqnarray}
	\sigma_{tW}^{theo} = 71.7 \pm 1.8 \mathrm{(scale)} \pm 3.4 \mathrm{(pdf)} \, . \nonumber
\end{eqnarray}
\setlength\belowdisplayskip{0.5pt}
\paragraph*{Summary}
A brief overview has been given of some recent total cross section 
measurements of \ttbar\ and single top production by ATLAS and CMS at the LHC, 
with unprecedented 
precision reached in several channels. Particular attention is paid to the 
\textit{in-situ} techniques in the ATLAS dilepton measurement that allows for the reduction of systematic 
uncertainties. A highlight from CMS is the first measurement of \ttbar\ production at 
13 TeV with selection targeting a $\tau$ lepton. Comparisons and combinations of the 
results from ATLAS and CMS help verify the results and improve the precision.





\begin{thebibliography}{99}


\bibitem{atlas}
ATLAS Collaboration, 2008 JINST 3 S08003.

\bibitem{cms}
CMS Collaboration, 2008 JINST 3 S08004.

\bibitem{atl_dilep}
ATLAS Collaboration, \href{https://link.springer.com/article/10.1140/epjc/s10052-020-7907-9}{Eur.Phys.J.C 80 (2020) 6, 528}.

\bibitem{blue1}
L. Lyons, D. Gibaut, P. Clifford, \href{https://doi.org/10.1016/0168-9002(88)90018-6}{Nucl. Instrum. Methods A 270, 110 (1988)}

\bibitem{blue2}
A. Valassi, \href{https://doi.org/10.1016/S0168-9002(03)00329-2}{Nucl. Instrum. Methods A 500, 391 (2003)}.


\bibitem{cms_dilep}
CMS Collaboration, \href{https://arxiv.org/pdf/1812.10505.pdf}{Eur.Phys.J.C 79 (2019) 5, 368}.

\bibitem{atl_ljets}
ATLAS Collaboration, \href{https://doi.org/10.1016/j.physletb.2020.135797}{Phys. Lett. B 810 (2020) 135797}.

\bibitem{cms_tau}
CMS Collaboration, \href{https://arxiv.org/pdf/1911.13204.pdf}{JHEP 02 (2020) 191}.


\bibitem{cms_tch}
CMS Collaboration, \href{https://arxiv.org/pdf/1812.10514.pdf}{Phys.Lett.B 800 (2020) 135042}.

\bibitem{combi}
ATLAS Collaboration and CMS Collaboration, \href{https://arxiv.org/pdf/1902.07158.pdf}{JHEP 05 (2019) 088}.

\bibitem{atl_tW}
ATLAS Collaboration, \href{https://arxiv.org/pdf/2007.01554.pdf}{2007.01554}, submitted to EPJC.

\bibitem{cms_tW}
CMS Collaboration, \href{https://link.springer.com/article/10.1007/JHEP10(2018)117}{JHEP 10 (2018) 117}.





\end{thebibliography}
\end{document}